\documentclass[aps,prl,twocolumn,amsmath,showpacs,amssymb,superscriptaddress,nofootinbib]{revtex4-1}

\usepackage{graphics,graphicx}
\usepackage{amsmath, amssymb}
\usepackage{multirow}
\usepackage{longtable}
\usepackage{color}

\usepackage[normalem]{ulem}  % \sout{old text} for strikeout

\begin{document}
\title{S-matrix analysis of the baryon electric charge correlation}
%\title{S-matrix analysis of the correlation of net baryon number with electric charge}
\date{\today}
\author{Pok Man Lo}
\affiliation{Institute of Theoretical Physics, University of Wroclaw,
PL-50204 Wroc\l aw, Poland}
\affiliation{Extreme Matter Institute EMMI, GSI,
Planckstr. 1, D-64291 Darmstadt, Germany}
\author{Bengt Friman}
\affiliation{GSI, Helmholzzentrum f\"{u}r Schwerionenforschung,
Planckstr. 1, D-64291 Darmstadt, Germany}
\author{Krzysztof Redlich}
\affiliation{Institute of Theoretical Physics, University of Wroclaw,
PL-50204 Wroc\l aw, Poland}
\affiliation{Extreme Matter Institute EMMI, GSI,
Planckstr. 1, D-64291 Darmstadt, Germany}
\author{Chihiro Sasaki}
\affiliation{Institute of Theoretical Physics, University of Wroclaw,
PL-50204 Wroc\l aw, Poland}

%\Large
%\preprint{YITP-12-91}
\begin{abstract}
We compute the correlation of the net baryon number with the electric charge ($\chi_{BQ}$) for an interacting hadron gas 
using the S-matrix formulation of statistical mechanics.
The observable $\chi_{BQ}$ is particularly sensitive to the details of the pion-nucleon interaction, 
which are consistently incorporated in the current scheme via the empirical scattering phase shifts.
Comparing to the recent lattice QCD studies in the $(2+1)$-flavor system,
we find that the natural implementation of interactions and the 
proper treatment of resonances in the S-matrix approach
lead to an improved description of the lattice data over that obtained in the hadron resonance gas model.
\end{abstract}

\maketitle

\vskip 0.2cm
\noindent
{\it Introduction.}-- Recent lattice QCD (LQCD) results on the equation of states and 
the fluctuations of conserved charges provide a very detailed description of the QCD thermal medium~\cite{l1,l2,l3,l4,l5}. 
In particular the local fluctuations of conserved charges can be probed by 
appropriate combinations of mixed susceptibilities. 
An accurate determination of these quantities is also needed to 
reliably extend the LQCD calculations to finite densities using the Taylor's expansion scheme~\cite{Bazavov:2017dus}.

Confinement dictates that hadrons, instead of quarks and gluons, fill the physical spectrum of QCD, 
while the spontaneous breaking of chiral symmetry makes pions exceptionally light due to their role as (pseudo-) Goldstone bosons.
We thus expect that at low temperatures the partition function can be effectively described by an interacting gas of 
low-mass hadrons such as pions, kaons, and nucleons.

A well-known effective approach which adopts the hadronic degrees of freedom in describing the thermodynamics
of strongly interacting matter is the hadron resonance gas (HRG) model.
This model assumes that resonance formation dominates the interactions of the confined phase, and as a first approximation, 
treats the resonances as an ideal gas.
The approach gives a satisfactory description of the particle yields measured in heavy ion collisions~\cite{h1,h2,h3,h4,h5,h6,h7,h8}, 
and is capable of providing an overall successful interpretation of LQCD results on bulk properties below the transition temperature~\cite{l1,l2,l3,l4,l5,l6,l7,l8}.

Nevertheless, the HRG model also makes some simplifying assumptions which are not necessarily consistent with the known hadron physics. 
Some of the problematic cases include the zero-width treatment of broad resonances~\cite{Weinhold:1997ig, Broniowski:2015oha,kappa} (e.g. the $\sigma$- and $\kappa$-meson), 
and the neglect of non-resonant contributions from both attractive and repulsive channels in computing the thermal observables~\cite{rho}.

\begin{figure}[!ht]
\includegraphics[width=3.355in]{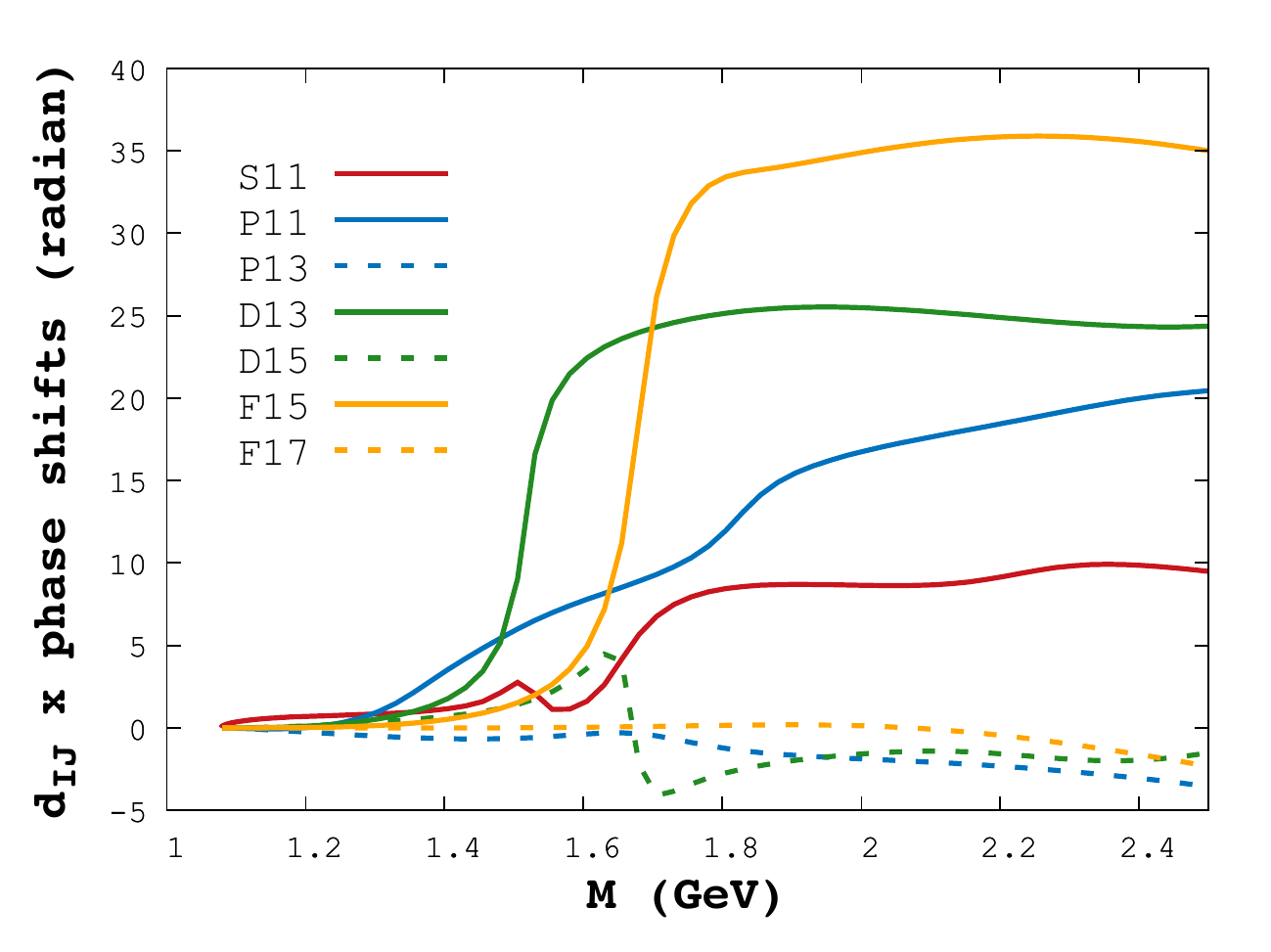}
\includegraphics[width=3.355in]{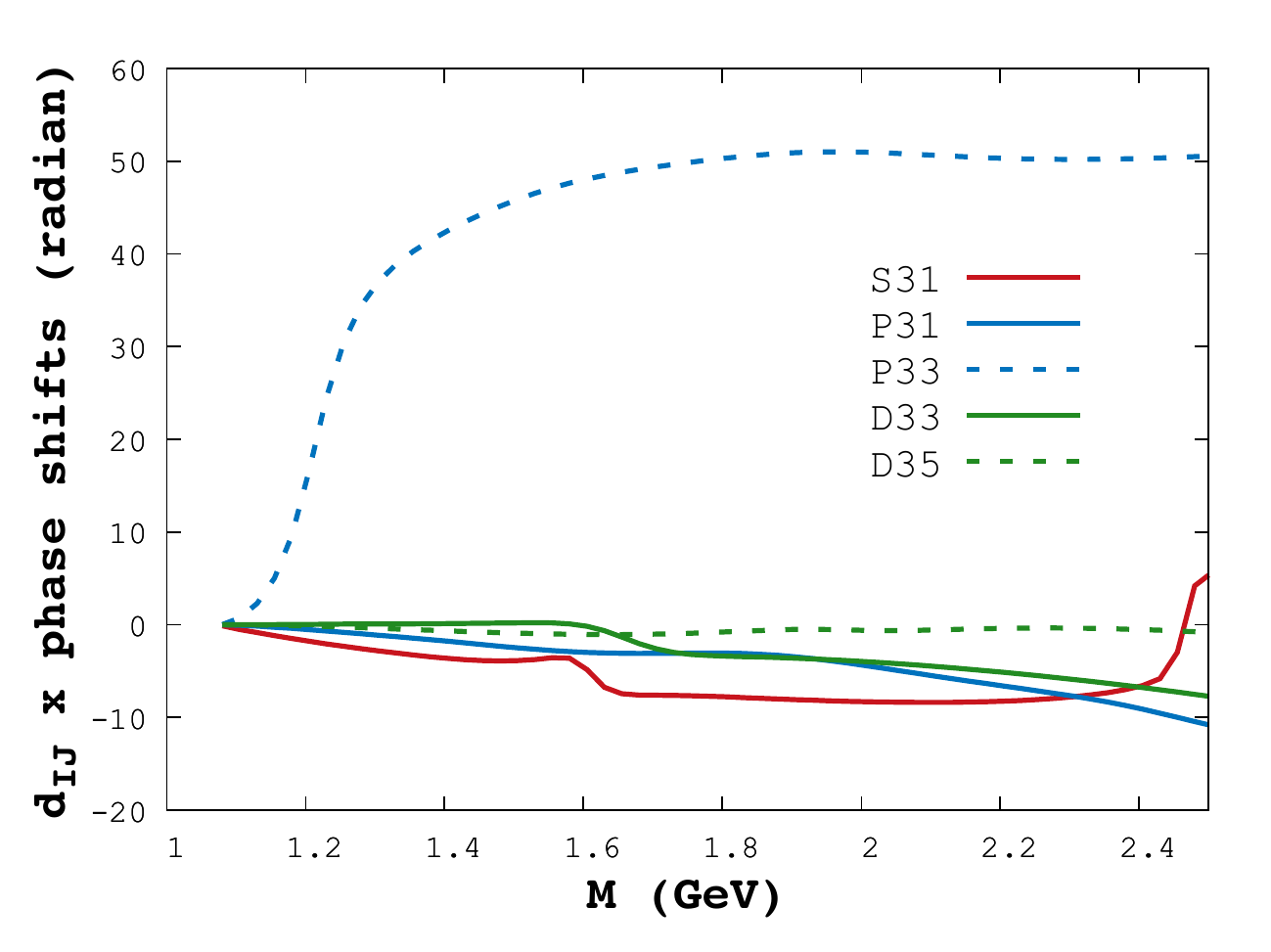}
  \caption{(color online). 
  $\pi N$ scattering phase shifts from SAID PW Analysis~\cite{Workman:2012hx}. 
  Shown in the figures are the major channels contributing to the observable $\chi_{BQ}$.}
\label{fig:fig1}
\end{figure}

Very precise information about the hadronic interactions has emerged from the impressive volume of experimental data~\cite{Olive:2016xmw}, carefully analyzed by theory such as chiral perturbation theory~\cite{Gasser:1983yg, Oller}, lattice QCD~\cite{Shepherd:2016dni}, effective hadron models~\cite{Rapp:1999ej} and potential models~\cite{Godfrey:1985xj,Barnes:1991em}. These studies have yet to be systematically included in the thermal models.

A promising approach to partly bridge this gap is the S-matrix formalism~\cite{Dashen:1969ep, Venugopalan:1992hy, Weinhold:1997ig, smat}. In this theoretical scheme, the two-body interactions of hadrons are included, via the scattering phase shifts, into an interacting density of states. This quantity is then folded into an integral over thermodynamic distribution functions, which then yields the interaction contribution to a particular thermodynamic observable. 

In this letter, we analyze the recent LQCD result~\cite{Bazavov:2017dus,Karsch:2017mvg,Karsch:2016yzt} 
on the baryon electric charge correlation $\chi_{BQ}$ using the S-matrix formalism. 
Unlike the common bulk quantities such as the pressure, this observable does not contain the large contribution from the purely mesonic channels. Furthermore, the contribution from the $\vert S \vert = 1$ strange baryons cancels in isospin symmetric systems.
This observable therefore demands a precise treatment of the nucleons and their interaction with pions. 
The latter is encoded in the comprehensive database of $\pi N$ scattering phase shifts which serve as an input for our study.

We show that a consistent treatment of the $\pi N$ interaction by the S-matrix formalism
leads to an improved description of lattice results up to a temperature $T\sim 160$ MeV, compared to that obtained in the HRG model.

\begin{table*}[htp!]
\begin{center}

\begin{tabular}{|c|c|c|c|c|}
			\hline
      \, P.W. (I=1/2) \, & \,	PDG resonances  \, 	& \multicolumn{2}{c|}{ \, channel VS total \, }	&  \, S-matrix VS HRG \,		\\  \hline
      \, $\,L_{2I,2J}\,(J^P)\,$             & $N^*$                  & \multicolumn{2}{|c|}{\, S-matrix	 ;      HRG\, } &   \\ \hline\hline
      $ \, S11\,(1/2-): \pi N + \eta N$\, & 1535, 1650          & \multicolumn{2}{|c|}{\, $0.139 + 0.114 \, (=0.253)$ ; 0.139\,}			      & \, $0.701 + 0.575 \, (=1.276)$ \,\\ \hline
      $ \, P11\,(1/2+)\,$	& 1440, 1710          & \multicolumn{2}{|c|}{0.201 ; 0.200}   				& 0.703					\\ \hline
      $ \, P13\,(3/2+)\,$	& 1720, (1900)         & \multicolumn{2}{|c|}{-0.046 ; 0.088}   			& -0.365					\\ \hline
      $ \, \underline{D13}\,(3/2-)\,$	& 1520, 1700, (1875)   & \multicolumn{2}{|c|}{0.329 ; 0.306} 	& 0.754					\\ \hline
      $ \, D15\,(5/2-)\,$	& 1675          & \multicolumn{2}{|c|}{0.035 ; 0.125}   				      & 0.196					\\ \hline
      $ \, F15\,(5/2+)\,$	& 1680          & \multicolumn{2}{|c|}{0.198 ; 0.118}   				      & 1.171					\\ \hline
      $ \, \forall I = 1/2 \,$	&               & \multicolumn{2}{|c|}{}   				              & 0.700					\\ \hline
\end{tabular}
    \caption{
      Comparison of the results on $\chi_{BQ}$ calculated in the S-matrix formalism and the HRG model in the major contributing channels for $I = 1/2$ at $T = 154 $ MeV.
      (The most dominant channel is underlined.)
      Entries in the column "channel VS total" correspond to the result computed within the given channel versus the sum of all available channels in $I=1/2$.
      The last column shows the ratio of $\chi_{BQ}$s computed within a given channel in the S-matrix approach and in the HRG model. 
      The last row shows the corresponding ratio for the sum of all channels.
      }
		\label{tab:1}
		\end{center}
\end{table*}

\begin{table*}[htp!]
\begin{center}

\begin{tabular}{|c|c|c|c|c|}
			\hline
      \, P.W. (I=3/2) \, & \,	PDG resonances  \, 	& \multicolumn{2}{c|}{ \, channel VS total \, }	&  \, S-matrix VS HRG \,		\\  \hline
      \, $\,L_{2I,2J}\,(J^P)\,$             & $\Delta$                  & \multicolumn{2}{|c|}{\, S-matrix	 ;      HRG\, } &   \\ \hline\hline
      $ \, S31\,(1/2-)\,$	& 1620          & \multicolumn{2}{|c|}{-0.110 ; 0.039}   				  & -1.841					\\ \hline
      $ \, P31\,(1/2+)\,$ & 1910          & \multicolumn{2}{|c|}{-0.048 ; 0.009}			      & -3.518					\\ \hline
      $ \, \underline{P33}\,(3/2+)\,$	& 1232, 1600, 1920   & \multicolumn{2}{|c|}{1.149 ; 0.829}   				& 0.911					\\ \hline
      $ \, D33\,(3/2-)\,$	& 1700   & \multicolumn{2}{|c|}{-0.004 ; 0.045}   				& -0.061					\\ \hline
      $ \, D35\,(5/2-)\,$	& 1930   & \multicolumn{2}{|c|}{-0.014 ; 0.019}   				& -0.502					\\ \hline
      $ \, F35\,(5/2+)\,$	& 1905   & \multicolumn{2}{|c|}{-0.003 ; 0.028}   				& -0.071					\\ \hline
      $ \, F37\,(7/2+)\,$	& 1950   & \multicolumn{2}{|c|}{0.026 ; 0.028}   				& 0.605					\\ \hline
      $ \, \forall I = 3/2 \,$	&               & \multicolumn{2}{|c|}{}   				                  & 0.657					\\ \hline
\end{tabular}
    \caption{Similar to the Table~\ref{tab:1} but for $I = 3/2$.}
		\label{tab:2}
		\end{center}
\end{table*}

\vskip 0.2cm
\noindent
{\it S-matrix treatment of the $\pi N$ system.}-- We first consider the thermodynamics of the $\pi N$ system at finite temperature and vanishing chemical potentials. 
In the S-matrix approach to statistical mechanics, the interaction contribution to the thermodynamic pressure from two-body scatterings
involves an integral over the invariant mass $M$~\cite{Weinhold:1997ig}, 

\begin{align}
\label{eqn:pressure}
  \begin{split}
    \Delta P_{int.} &= \frac{T}{V} (\ln Z)_{int.} \\
   &\approx \sum_{I_z;B=-1,1} d_{j} \times T \int_{m_{th}}^\infty d M \int \frac{d^3 p}{(2 \pi)^3} \, \frac{1}{\pi} \frac{d \delta^I_j}{d M}   \\
   &\times \left[ \ln \, (1 + e^{-\beta \sqrt{p^2+M^2}}) \right]. 
  \end{split}
\end{align}

\noindent where $\delta^I_j$ is the scattering phase shift for a given isospin-spin channel and $d_{j}$ is the degeneracy factor for spin $j$. 
We have made the sum over the isospin states explicit. In addition, the antiparticle contribution is implemented via the sum over the baryon number $B$, with $B = 1(-1)$ for baryons (antibaryons). An analogous expression can be derived for the susceptibilities. Specifically for $\chi_{BQ}$ the expression reads

\begin{align}
\label{eqn:sus}
  \begin{split}
    \Delta \chi_{BQ} & \approx \sum_{I_z;B=-1,1} d_{j} \times  B \times Q \times \frac{1}{T} \\
    &\times \int_{m_{th}}^\infty d M \int \frac{d^3 p}{(2 \pi)^3} \, \frac{1}{\pi} \frac{d \delta^I_j}{d M} \\
    &\times \left[ \frac{e^{-\beta \sqrt{p^2+M^2}}}{(1 + e^{-\beta \sqrt{p^2+M^2}})^2} \right].
  \end{split}
\end{align}

\noindent The electric charge $Q$ of a hadron composed of light quarks can be related to the baryon number $B$, the strangeness $S$, 
and the $z$-component of the isospin $I_z$, via the Gellmann-Nishijima formula~\cite{Olive:2016xmw}

\begin{align}
  \label{eqn:GN_formula}
  Q = I_z + \frac{1}{2} \left( B + S \right).
\end{align}

\noindent From Eq.~\eqref{eqn:sus} and \eqref{eqn:GN_formula}, we deduce that the $\Lambda$- and the $\Sigma$-baryons, i.e. the $\vert S \vert  = 1 $ hyperons, do not contribute to $\chi_{BQ}$. This is due to the chargeless ($Q=0$) nature of the former and the explicit cancellation between the $Q=-1$ and $Q=1$ states in the isospin triplet for the latter. This is generally true for an isospin-symmetric system, regardless of the details in treating the resonances.

It is therefore convenient to separate the observable $\chi_{BQ}$ into two parts: the contribution from free nucleons and $\vert S \vert=2, 3$ baryons; and the contribution from $\pi N$ interaction. We treat the former as a free gas and compute the latter using Eq.~\eqref{eqn:sus}. 

The key quantity in the S-matrix formalism is the effective density of states

\begin{align}
\frac{1}{\pi} \frac{d \delta^I_j}{d M}, 
\label{eq:dos}
\end{align}

\noindent which can be derived from the scattering phase shifts and thus contains the dynamics of the $\pi N$ interaction.

In this study, 15 partial waves (PWs) for each of the isospin channels ($I=1/2, 3/2$) from the SAID PW Analysis~\cite{Workman:2012hx} have been included to compute the effective spectral function. 
The major contributing channels are reproduced in Fig.~\ref{fig:fig1}.

Using the empirical phase shifts, it is straightforward to compute their contributions to $\chi_{BQ}$ numerically. 
The result turns out to be dominated by a few channels at low angular momenta $L$: D13, S11, P11, and F15 for the case of $N^*$ resonances; 
P33 and S31 for the case of $\Delta$. 
Details of the contributions from major channels for $I=1/2$ and $I=3/2$ at $T = 154$ MeV 
are summarized in tables~\ref{tab:1} and ~\ref{tab:2}.
Comparisons to the HRG model are also included.

\begin{figure*}[ht!]
 \includegraphics[width=0.49\textwidth]{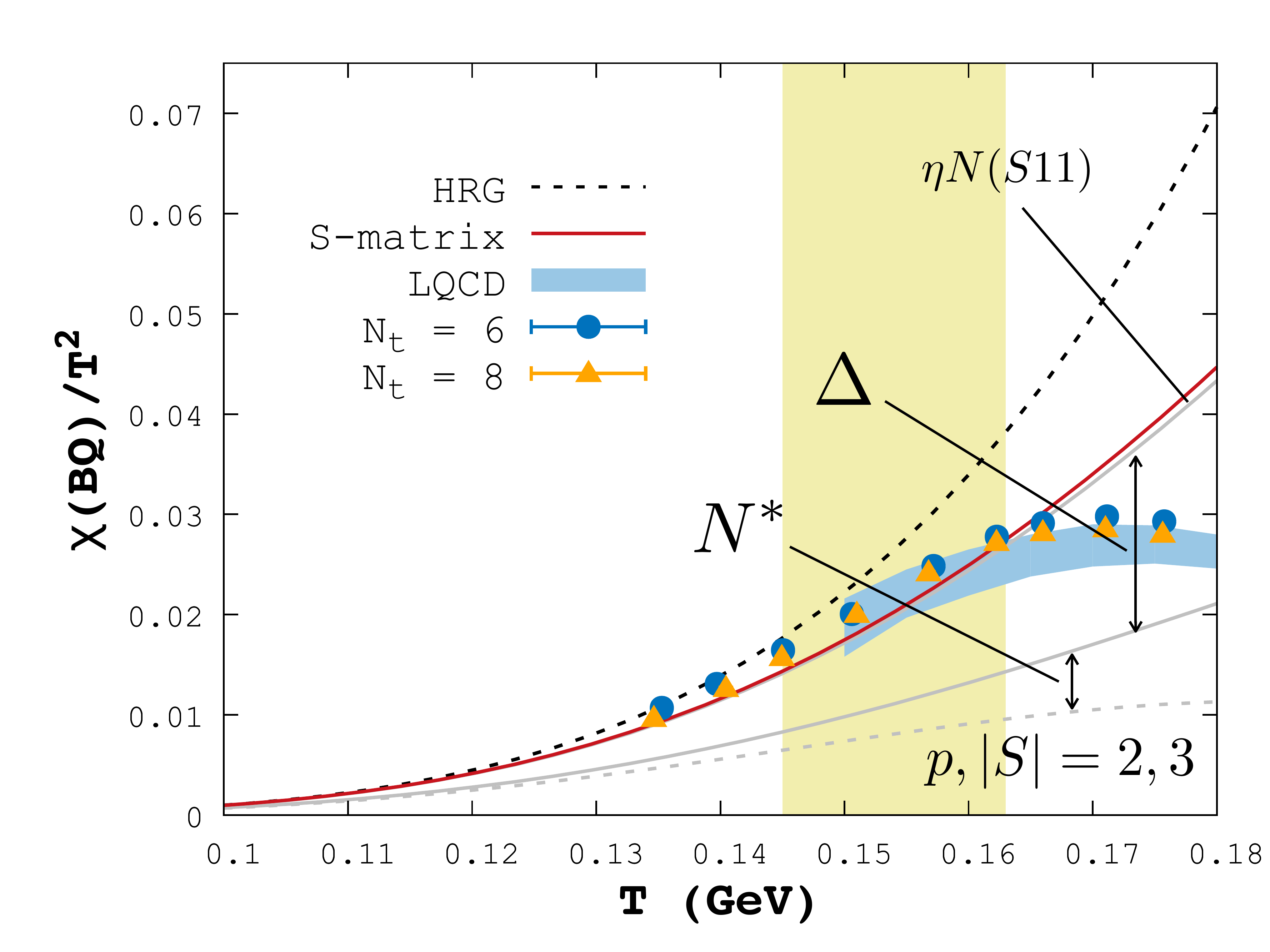}
 \includegraphics[width=0.49\textwidth]{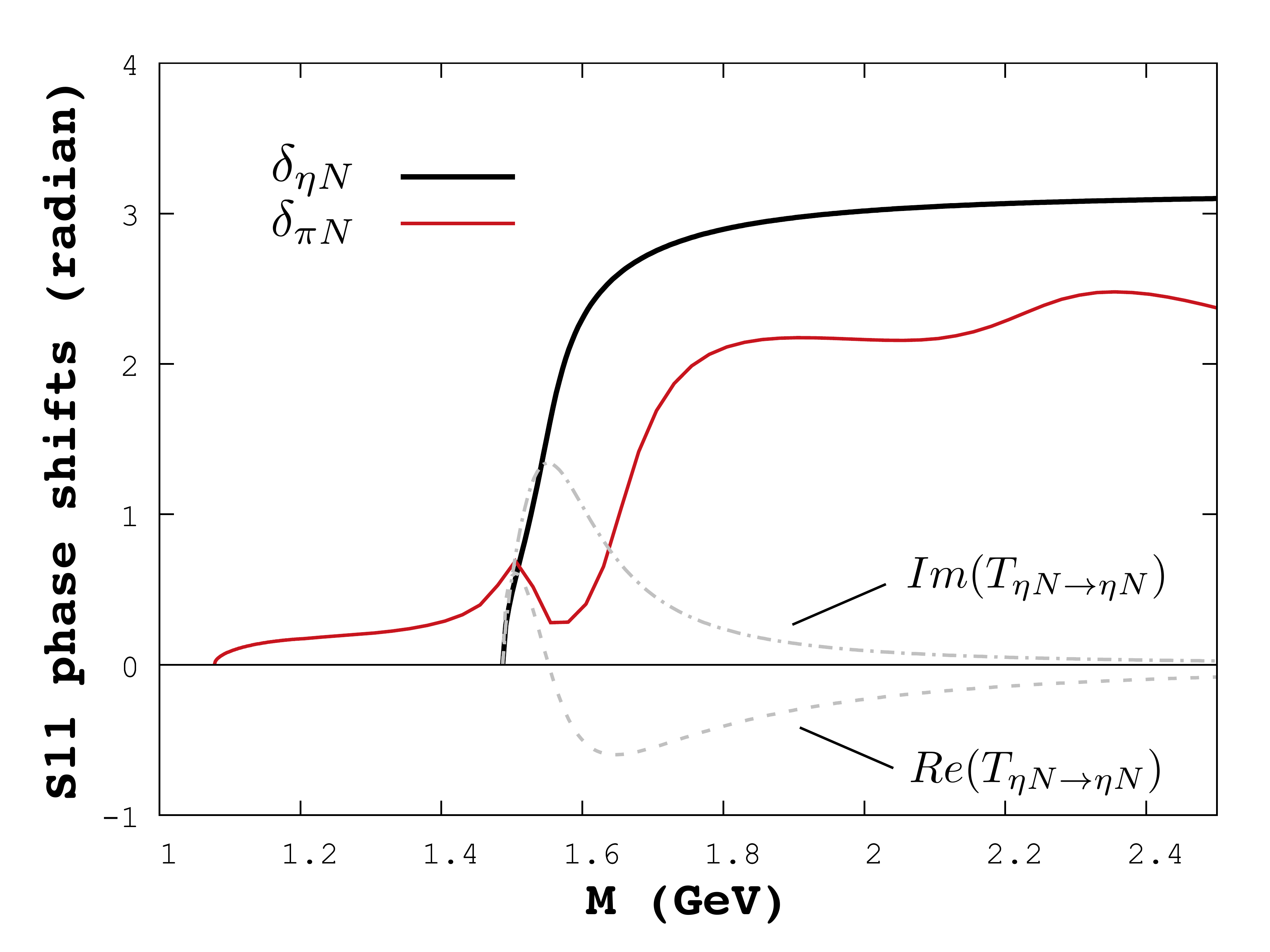}
  \caption{Left: the susceptibility $\chi_{BQ}$, scaled to dimensionless units, from LQCD compared to the predictions by various theoretical approaches. In this work the S-matrix treatment is restricted to the $\pi N$ sector with an additional S11 channel of $\eta N$. For the former a total of $30$ PW channels (15 each for $I = 1/2, 3/2$) are included.
  The yellow band indicates the range of pseudo-critical temperature $T_c = 154 \pm 9\,  {\rm MeV}$. Right: Phase shifts in the S11 channel: $\delta_{\pi N}$ provided by Ref.~\cite{Workman:2012hx} and $\delta_{\eta N}$ extracted from Ref.~\cite{Green:2004tj}. Also shown are the T-matrix elements computed from Eqs.~\eqref{eq:t_matrix} and~\eqref{eq:ere}.
  }
\label{fig:fig2}
\end{figure*}

Summing over the contributions of the available $\pi N$ phase shifts, we obtain the S-matrix result on $\chi_{BQ}$. This is shown in Fig.~\ref{fig:fig2} (left), together with the HRG model and LQCD results~\cite{Karsch:2016yzt}. 
The dominant contribution comes from the $I = 3/2$ sector but the $I = 1/2$ states also play an important role. The overall result obtained in the S-matrix approach is substantially lower than that of the HRG model, approaching the tentative LQCD values in the chiral crossover region. 
At $T = 154$ MeV, the overall suppression of the interaction contribution in the S-matrix approach, compared to the HRG model, is around $30 \, \%$. 
At temperatures beyond the crossover region, approaches based on the hadronic degrees of freedom are expected to break down and 
can no longer provide a reliable description of the LQCD result.

The source of the improvement in the quantitative description of the LQCD result within the S-matrix approach is twofold. First, the inclusion of non-resonant, often purely repulsive, channels yields an important contribution at low invariant masses~\cite{kappa,exvol}. Second, a consistent treatment of the interactions is pivotal in channels with broad resonances. For such a resonance, the thermal contribution can be significantly reduced relative to the HRG prediction owing to the fact that a substantial part of the effective density of states (\ref{eq:dos}) is found at large masses, which are suppressed by the Fermi-Dirac or Bose-Einstein factors. This effect is illustrated for the case of $\sigma$- and $\kappa$-mesons in Refs.~\cite{Broniowski:2015oha,kappa}. 

Naturally, the above corrections to the effective density of states should be taken into account for all the thermodynamic quantities. Nevertheless, it is in the
observable $\chi_{BQ}$ that such a precision calculation becomes increasingly important and the effect becomes clearly visible.

\vskip 0.2cm
\noindent
{\it Effects of inelasticity.}--The S-matrix analysis presented so far is, strictly speaking, applicable only to the case of elastic scattering. As a first estimate, we
restrict our investigation to the two-body subspace, where the first inelastic channel opens at the $\eta$-production threshold, i.e. $M = m_\eta + m_N \approx 1.5 \, {\rm GeV}$.
Therefore, in addition to $ \pi + N \rightarrow \pi + N $, the scattering processes $ \pi + N \rightarrow \eta + N $ 
and $ \eta + N \rightarrow \eta + N $ need to be taken into account in constructing the effective density of states~\eqref{eq:dos}. To properly account for the inelastic process $\pi N \rightarrow \pi\pi N$, requires an extension of the scheme employed here. We defer this to a future publication. 

The S-matrix for the coupled-channel problem can be expressed in terms of two scattering phase shifts ($\delta_{\pi N}$, $\delta_{\eta N}$) 
and an inelasticity parameter $\alpha$ via~\cite{K-matrix, smat}, as

\begin{align}
  S = \left( \begin{array}{cc}
    \alpha \, e^{2  i  \delta_{\pi N}}  &   i \sqrt{1-\alpha^2} \, e^{i (\delta_{\pi N} + \delta_{\eta N})} \\
    i \sqrt{1-\alpha^2} \, e^{i  (\delta_{\pi N} + \delta_{\eta N})}   &  \alpha \, e^{2  i  \delta_{\eta N}}
  \end{array} \right).
\end{align}

\noindent The modification required of the current formulation to treat this case is by replacing $\delta_{\pi N} \rightarrow \mathcal{Q}(M)$~\cite{smat}

\begin{align}
  \begin{split}
    \mathcal{Q}(M) &\equiv \frac{1}{2} \, \rm{Im} \left( {\rm tr} \ln \, S \right) \\
    &= \frac{1}{2} \, \rm{Im} \left( \ln \, {\rm det} \,[S] \right)\\
    &= \delta_{\pi N} + \delta_{\eta N}.
  \end{split}
\end{align}

\noindent Note that the expression is independent of the inelasticity parameter $\alpha$. Hence all that is required to compute thermodynamic quantities is the additional phase shift $\delta_{\eta N}$.

The status of the empirical data on $\delta_{\eta N}$ is unfortunately far more uncertain than the $\pi N$ case. 
Robust modeling of the scattering amplitude~\cite{Batinic:1995kr, Batinic:1997gk, Sauermann:1994pu, DeutschSauermann:1997mt, Green:2004tj} is only available for a few PW channels.
In this study, we focus on the S11 channel and make use of the effective range expansion scheme in Ref.~\cite{Green:2004tj} to extract the required phase shift.

We recapitulate briefly how to extract the phase shift from the relevant T-matrix element. In our notation, the T-matrix element describing the scattering
process $\eta N \rightarrow \eta N$ is

\begin{align}
  \begin{split}
    \label{eq:t_matrix}
    T_{\eta N \rightarrow \eta N} &= (\alpha \, e^{2  i  \delta_{\eta N}} - 1)/i \\
    &= \frac{2}{\mathcal{C}_{\eta N} - i }.
  \end{split}
\end{align}

\noindent The complex function $\mathcal{C}_{\eta N}$ admits an effective range expansion as follows:

\begin{align}
  \label{eq:ere}
  \mathcal{C}_{\eta N} = \frac{1}{q a} + \frac{1}{2} r_0 q + s q^3 + \dots,
\end{align}

\noindent where $q$ is the momentum of the particle in the center-of-mass frame, related to $M$ via

\begin{align}
  q = \frac{M}{2} \, \sqrt{ 1 - \frac{(m_N+m_\eta)^2}{M^2}} \, \sqrt{ 1 - \frac{(m_N-m_\eta)^2}{M^2}}.
\end{align}

\noindent Using the parameters given in Ref.~\cite{Green:2004tj} 

\begin{align}
  \begin{split}
    a({\rm fm}) &= 0.91(6) + i \, 0.27(2) \\
    r_0({\rm fm}) &= -1.33(15) - i \, 0.30(2) \\
    s({\rm fm^3}) &= -0.15(1) - i \, 0.04(1), \\
  \end{split}
\end{align}

\noindent the phase shift $\delta_{\eta N}$ can be extracted from

\begin{align}
  \delta_{\eta N} = \frac{1}{2} \, {\rm Im} \, \ln \left( 1 + i \, T_{\eta N \rightarrow \eta N} \right). 
\end{align}

\noindent The results are shown in Fig.~\ref{fig:fig2} (right). 
Phase shifts in other PWs are not as well-constrained~\cite{Batinic:1995kr} 
but can similarly be included in the analysis once the situation improves.

The contribution to $\chi_{BQ}$ from $\delta_{\eta N}$ can be readily computed. 
This adds to the previous result of $\pi N$ scatterings and give the total S-matrix result of this work (red line) in Fig.~\ref{fig:fig2}.
As seen in the figure, the contribution from inelasticity to the observable remains quite small~\footnote{Though the contribution from $\eta N$ is small compared to the overall contribution, it is definitely not small compared to the other contribution in the S11 channel. The S-matrix treatment of the interaction may be interesting for studying the $\eta$-production physics~\cite{Batinic:1995kr}.}. 
This is mainly due to fact that the Boltzmann suppression of the large invariant mass $M$ contribution is strong at low temperatures. 
In fact, we have checked that up to $T \approx 0.16 \, {\rm GeV}$, over $90\%$ of the value of $\chi_{BQ}$ comes from the elastic part of the spectrum ($M \leq m_\eta + m_N$). 
This is reassuring since the large-$M$ region of the spectrum is generally poorly known.
It also stresses the importance of an accurate treatment of the low invariant mass region.

At larger temperatures, effective models of QCD based on hadronic degrees of freedom will eventually break down. A description of the thermodynamics of strongly interacting matter at these temperatures
require a mechanism for hadron dissociation and the implementation of explicit quark and gluon degrees of freedom.
Moreover, as the suppression due the Boltzmann factor is less effective at higher temperatures, a proper treatment of the details of the high-mass spectrum becomes necessary.
The calculation presented in this work is therefore approximate. Nevertheless, it serves as a baseline, where the known vacuum physics is implemented via a consistent treatment of two-body interactions in studies of the thermodynamics of strongly interacting matter in the hadronic phase.

\vskip 0.2cm
\noindent
{\it Conclusions.}--We have analyzed the recent lattice QCD result on the baryon electric charge correlation $\chi_{BQ}$ 
at vanishing chemical potentials within the S-matrix formalism. 
The observable $\chi_{BQ}$ is particularly sensitive to the interaction between pions and nucleons.
In the current framework, the hadronic forces are consistently incorporated in the effective density of states via the empirical scattering phase shifts.
Specifically, purely repulsive channels and the relevant resonances are naturally included in the calculation.
This yields an improved description of the lattice result over that of the hadron resonance gas model.

%=========================

The calculation presented in this work serves as a baseline in which 
the known vacuum physics is consistently implemented in studying the thermodynamics 
under a virial expansion scheme up to the second order.
This can be the necessary first step to properly incorporating further in-medium effects~\cite{Rapp:1995fv, Lacroix:2014gsa, Aarts:2017rrl}.

A further study could extend the S-matrix analysis to other lattice observables such as the
ratios $\chi_{BQ}/\chi_{BS}$ and $\chi_{BQ}/\chi_{BB}$. 
Unlike the observable $\chi_{BQ}$, the susceptibilities $\chi_{BB}$ and $\chi_{BS}$ receive contributions 
from all the baryonic channels. 
The latter is particularly sensitive to the strangeness content of the medium, including the $\vert S \vert = 1$ 
hyperons~\cite{Bazavov:2014xya, hag}.
An S-matrix study of these correlations is challenging
since the available data is not sufficient to allow a high-quality partial-wave analysis.
However, the improved agreement with lattice data obtained in this work provides a motivation for further studies in this direction. 

%========================

{\it Acknowledgments.}--We thank Frithjof Karsch for fruitful discussions on the recent LQCD results.
We are also grateful to Peter Braun-Munzinger, Pasi Huovinen and Kenji Morita for constructive comments.
This work was partly supported by the Polish National Science Center (NCN), under Maestro Grant No. DEC-2013/10/A/ST2/00106 
and by the ExtreMe Matter Institute EMMI at the GSI Helmholtzzentrum fuer Schwerionenphysik, Darmstadt, Germany.

\end{document}